\documentclass{aastex61}

\usepackage{color}
\usepackage{url}
\usepackage{epstopdf}
\usepackage{multirow}
\usepackage{amsmath}
\usepackage{amstext}
\usepackage{natbib}
\usepackage{longtable}

\received{June 30, 2017}
\revised{\today}
\accepted{\today}

\submitjournal{AAS Journals}

\shorttitle{Optical and NIR Polarimetry of C/2013 US10}
\shortauthors{Yuna Grace Kwon et al.}

\begin{document}

\title{Optical and Near-Infrared Polarimetry of Non-Periodic Comet C/2013 US10 (Catalina)}


\correspondingauthor{Masateru Ishiguro}
\email{ishiguro@astro.snu.ac.kr}

\author{Yuna Grace Kwon} 
\author{Masateru Ishiguro}
\affil{Department of Physics and Astronomy, Seoul National University,\\1
Gwanak, Seoul 08826, Republic of Korea}

\author{Daisuke Kuroda}
\affil{Okayama Astrophysical Observatory, National Astronomical
Observatory of Japan, \\ 3037-5 Honjo, Kamogata, Asakuchi, Okayama, 719-0232, Japan}

\author{Hidekazu Hanayama}
\affil{Ishigakijima Astronomical Observatory, National Astronomical
Observatory of Japan, \\ 1024-1 Arakawa, Ishigaki, Okinawa 907-0024, Japan}

\author{Koji S. Kawabata} 
\affil{Hiroshima Astrophysical Science Center, Hiroshima University, \\
Kagamiyama 1-3-1, Higashi-Hiroshima 739-8526, Japan}

\author{Hiroshi Akitaya} 
\affil{Center for Astronomy, Ibaraki University, \\
2-1-1 Bunkyo, Mito, Ibaraki 310-8512, Japan}

\author{Tatsuya Nakaoka}
\affil{Hiroshima Astrophysical Science Center, Hiroshima University, \\
Kagamiyama 1-3-1, Higashi-Hiroshima 739-8526, Japan}

\author{Ryosuke Itoh}
\affil{Department of Physics, Tokyo Institute of Technology,
\\Meguro, Tokyo 152-8551, Japan}

\author{Hiroyuki Toda} 
\author{Kenshi Yanagisawa}
\affil{Okayama Astrophysical Observatory, National Astronomical
Observatory of Japan, \\ Asakuchi, Okayama 719-0232, Japan}

\author{Myung Gyoon Lee}
\affil{Department of Physics and Astronomy, Seoul National University,\\
Gwanak, Seoul 08826, Republic of Korea}

\author{Kouji Ohta}
\affil{Department of Astronomy, Kyoto University, Kyoto 606-8502, Japan}

\author{Michitoshi Yoshida}
\affil{Subaru Telescope, National Astronomical Observatory of Japan,\\
Hilo, HI 96720, USA}

\author{Nobuyuki Kawai}
\affil{Department of Physics, Tokyo Institute of Technology,
\\Meguro, Tokyo 152-8551, Japan}

\author{Jun-Ichi Watanabe}
\affil{National Astronomical Observatory, Mitaka, Tokyo 181-8588, Japan}

\begin{abstract}
We present an optical and near-infrared (hereafter NIR) polarimetric study of a comet C/2013 US10 (Catalina) observed on UT 2015 December 17--18 at phase angles of $\alpha$=52.1\arcdeg -- 53.1\arcdeg. Additionally, we obtained an optical spectrum and multi-band images to examine the influence of gas emission. We find that the observed optical signals are significantly influenced by gas emission, that is, the gas-to-total intensity ratio varies from 5 to 30 \% in the $R_{\rm C}$ and 3 to 18 \% in the $I_{\rm C}$ bands, depending on the position in the coma. We derive the `gas-free dust polarization degrees' of 13.8$\pm$1.0 \% in the $R_{\rm C}$ and 12.5$\pm$1.1 \% in the $I_{\rm C}$ bands and a gray polarimetric color, i.e., --8.7$\pm$9.9 \% \micron$^{-1}$ in optical and 1.6$\pm$0.9 \% \micron$^{-1}$ in NIR. The increments of polarization obtained from the gas correction show that the polarimetric properties of the dust in this low-polarization comet are not different from those in high-polarization comets. In this process, the cometocentric distance dependence of polarization has disappeared. We also find that the $R_{\rm C}$-band polarization degree of the southeast dust tail, which consists of large dust particles (100 \micron -- 1 mm), is similar to that in the outer coma where small and large ones are mixed. Our study confirms that the dichotomy of cometary polarization does not result from the difference of dust properties, but from depolarizing gas contamination. This conclusion can provide a strong support for similarity in origin of comets.
\end{abstract}


\section{Introduction}

Linear polarization of cometary dust particles has been investigated to determine the physical properties, such as sizes, compositions, and porosities \citep[see, e.g.,][]{{Kiselev et al.2015}}. Observationally, it is known that the degree of linear polarization of cometary dust ($P$) exhibits a strong dependence on the phase angle ($\alpha$, angle of Sun-comet-observer). It exhibits a bell-shaped curve: a shallow negative branch at 0\arcdeg $<\alpha <$ 20\arcdeg, an inversion angle at $\alpha_{\rm 0}$ $\sim$ 25\arcdeg, a quasi-linear increase over $\alpha$=25\arcdeg -- 60\arcdeg, and a maximum polarization degree of $P_{\rm max}\sim$10--30 \% at $\alpha \sim 90$\arcdeg\ \citep{Chernova et al.1993,Levasseur-Regourd et al.1996,Hadamcik et al.2003a,Hadamcik et al.2003b,Kolokolova et al.2004}. In addition, $P$ exhibits a moderate wavelength dependence.

\citet{Chernova et al.1993} and \citet{Levasseur-Regourd et al.1996} noted an existence of the dichotomy of polarimetric phase curve of comets at $\alpha\gtrsim45\arcdeg$. \citet{Chernova et al.1993} suggested that comets in the high-polarization class tend to be dust rich, whereas comets in the low-polarization class could be gas rich. Later, it was found that C/1995 O1 (Hale-Bopp) could not be classified into either of these classes, as it exhibits the highest polarization at a given phase angle \citep{Hadamcik et al.1997,Hadamcik et al.2003b}. This observational evidence indicated a need for further research mainly from two points of view: the polarimetric classes reflect inherent physical differences in the bulk properties of cometary dust particles \citep{Levasseur-Regourd et al.1996} and/or different degrees of contamination associated with gas emissions \citep{Kiselev et al.2004,Jockers et al.2005}. \citet{Chernova et al.1993} noted that depolarization of gas emission is indispensable even in data taken with narrow-band optical filters, which are less susceptive to gas emission, for studying the scattered light originating from dust particles. Thus, it is important to develop a technique to discriminate a signal of gas emission from the observed one to derive a gas-free $P$ of dust particles.

In this regard, the recent apparition of a bright (total apparent visual magnitude of $\sim$7 mag) comet, C/2013 US10 (Catalina) (hereafter C/2013 US10), provided a further opportunity to conduct polarimetric observations at  phase angles of $\alpha$=52.1\arcdeg -- 53.1\arcdeg\ and test a new technique to extract a dust signal from the observed data. The phase angle was moderately large, enabling us to distinguish the above polarimetric classes. By conducting complementary multi-band imaging and spectroscopic observations, we attempted to discriminate gas and dust signals. We derived the gas intensity rates as fractions of the total observed intensity in standard optical filters and estimated $P$ for the dust continuum. We describe our observations and the data analyses in Section 2 and the observational results in Section 3. We discuss the observed results in Section 4 and summarize the study in Section 5.

\section{Observations} 
\subsection{Observations with Ground-Based Telescopes}
We utilized  three ground-based telescopes that constitute a portion of the Optical and Infrared Synergetic Telescopes for Education and Research (OISTER) inter-university observation network in Japan: the 150-cm Kanata telescope at the Higashi-Hiroshima Observatory (hereafter, HHO), the 50-cm telescope at the Okayama Astrophysical Observatory (OAO), and the 105-cm Murikabushi telescope at the Ishigakijima Astronomical Observatory (IAO). A summary of these observations is given in Table 1. Detailed information about the observations at each observatory is presented below.

\begin{deluxetable}{ccccccccc}[h!]
\centering
\tablecaption{Journal of  observations \label{Information}}
\tablehead{
\multirow{2}{*}{UT date} & \multirow{2}{*}{Telescope/Instrument} & \multirow{2}{*}{Mode} & \multirow{2}{*}{Filter} & \multirow{2}{*}{Exptime} & $N$ & $r_{\rm h}$ & $\Delta$ & $\alpha$ \\
 & & & & & (1) & (2) & (3) & (4)
}
\startdata
2015/11/15.72  &  Perihelion  & \ldots &      \ldots        & \ldots & \ldots & 0.82 & \ldots & \ldots\\
\\
2015/12/17.79 & HHO / HONIR & Impol & $R_{\rm C}$ & 30 & 38 & 1.02 & 1.19 & 52.1\\
2015/12/17.79 & HHO  /HONIR & Impol & $K_{\rm S}$ & 15 & 47 & 1.02 & 1.19 & 52.1\\
2015/12/17.83 & HHO / HONIR & Impol & $I_{\rm C}$ &  30 & 22 & 1.02 & 1.19 & 52.1\\
2015/12/17.84 & OAO / MITSuME & Image & {\sl g}$'$, $R_{\rm C}$, $I_{\rm C}$ & 60 & 27 & 1.02 & 1.19 & 52.1\\
2015/12/18.83 & OAO / MITSuME & Image & {\sl g}$'$, $R_{\rm C}$, $I_{\rm C}$ & 60 & 27 & 1.03 & 1.17 & 52.7\\
2015/12/18.84 & HHO / HONIR & Impol & $R_{\rm C}$ & 30 &16 & 1.03 & 1.17 & 52.7\\
2015/12/18.84 & HHO / HONIR & Impol & $K_{\rm S}$ & 15 & 17 & 1.03 & 1.17 & 52.7\\
\multirow{ 2}{*}{2015/12/18.88} & \multirow{ 2}{*}{HHO / HONIR} & \multirow{ 2}{*}{Spec}  & OPT-058$^{*1}$ &\multirow{ 2}{*}{125} & \multirow{ 2}{*}{4} & \multirow{ 2}{*}{1.03} & \multirow{ 2}{*}{1.17} & \multirow{ 2}{*}{52.7}\\
& & & IR-1.33$^{*2}$ & & & & \\
2015/12/19.79 & OAO / MITSuME & Image & {\sl g}$'$, $R_{\rm C}$, I$_{\rm C}$ & 60 & 107 & 1.04 & 1.15 & 53.1\\
2015/12/19.81 & IAO / MITSuME & Image & {\sl g}$'$, $R_{\rm C}$, I$_{\rm C}$ & 60 & 57 & 1.04 & 1.15 & 53.1 \\
\enddata
\tablecomments{The top header shows the following quantities: (1) the number of the data sets, (2) the median heliocentric distance in au, (3) the median geocentric distance in au, and (4) the median phase angle in degree.~`Mode' stands for the observational mode: Image (imaging), Impol (imaging polarimetry), or Spec (spectroscopy). We used the web-based JPL Horizon system\footnote{http://ssd.jpl.nasa.gov/?horizons} to obtain these ephemerides. Each abbreviation for the telescopes denotes HHO (150-cm Kanata telescope at the Higashi-Hiroshima Observatory), OAO (50-cm telescope at the Okayama Astrophysical Observatory), and IAO (105-cm Murikabushi telescope at the Ishigakijima Astronomical Observatory). Exptime denotes the individual exposure time.}
\tablenotetext{*1}{OPT-058 denotes the Optical-Arm filter, whose wavelength coverage is 0.58\micron\ -- 1.00\micron\ \citep{Akitaya et al.2014}.}
\tablenotetext{*2}{IR-1.33 denotes the IR-Arm filter, whose wavelength coverage is 1.33\micron\ -- 2.40\micron\ \citep{Akitaya et al.2014}.}
\end{deluxetable}

\subsubsection{HHO Observations}
We performed polarimetric and spectroscopic observations of C/2013 US10 with the Hiroshima Optical and Near-InfraRed camera (HONIR) attached to the Cassegrain focus of the 150-cm Kanata telescope at the HHO (34\arcdeg 22\arcmin 39\arcsec N, 132\arcdeg 46\arcmin 36\arcsec E, 503 m), Hiroshima prefecture, Japan, on UT 2015 December 17--18. This instrument enables one to obtain imaging, spectroscopic and polarimetric data in both optical and NIR bands simultaneously \citep{Akitaya et al.2014}. We conducted the polarimetry observations with an optical CCD (2048 $\times$ 4096 pixels with a 15 \micron\ pixel pitch) covering the $R_{\rm C}$ and $I_{\rm C}$ bands and with the VIRGO-2k HgCdTe array (2048 $\times$ 2048 pixels with a 20 \micron\ pixel pitch) covering the $K_{\rm S}$ band. Each detector covers the same field-of-view (FOV) of 10\arcmin $\times$ 10\arcmin\ with the same pixel resolution of 0.29\arcsec. In imaging polarimetry mode, HONIR is equipped with a rotatable super-achromatic half-wave plate (HWP), a LiYF$_4$ (YLF) Wollaston prism, and a focal mask to avoid the overlap of images in ordinary and extraordinary rays. We chose exposure times of 30 and 15 seconds for the optical and NIR polarimetric observations, respectively, at each HWP angle (in the sequence of 0\arcdeg, 45\arcdeg, 22.5\arcdeg, and 67.5\arcdeg\ position angles to the fiducial point). In the spectroscopy mode, we employed a focal slit mask made of aluminum alloy with a 2.2\arcsec ~(0.2 mm)-wide slit, yielding R~(= $\lambda$/$\Delta$$\lambda$) $\sim$ 350, for both optical and NIR wavelengths. We set the exposure time of 125 seconds for the spectroscopic observation. Because of unavailability of the comet tracking mode, the telescope was operated in the sidereal tracking mode. The average seeing size was $\sim$ 2\arcsec\ on these nights. The elongation of the comet, which depends on the exposure time,  was smaller than that of the seeing disk size in the imaging polarimetric data (1.2\arcsec) but slightly longer than the seeing size in the spectroscopic data (4.8\arcsec). We accordingly forgo a discussion about fine-scale structure smaller than 5\arcsec\ (4300 km) in the following sections.

\subsubsection{OAO Observations}
Simultaneous imaging observations with the HHO were performed using the OAO 50-cm telescope atop Mt. Chikurinji (34\arcdeg 34\arcmin 33\arcsec N, 133\arcdeg 35\arcmin 36\arcsec E, 360 m), Okayama prefecture, Japan, on UT 2015 December 17--19. This telescope is designed to perform rapid follow-up observations of gamma-ray bursts and is useful for observations of transient solar system objects such as comets. We exploited the Multicolor Imaging Telescopes for Survey and Monstrous Explosions (MITSuME), which consists of three Alta U-6 cameras with 1024 $\times$ 1024 pixel CCDs with a 24 \micron\ pixel pitch each, to take images in three bands ({\sl g}$'$, $R_{\rm C}$, and $I_{\rm C}$ bands) simultaneously. The instrument is attached to the Cassegrain focus of the telescope, covering a FOV of 26\arcmin $\times$ 26\arcmin\ with a pixel scale of 1.5\arcsec\ \citep{Kotani et al.2005,Yanagisawa et al.2010}. The average FWHM on these nights was $\sim$ 2--3\arcsec. We applied the comet tracking mode of this telescope.

\subsubsection{IAO Observation}
In addition to the OAO, we obtained simultaneous imaging data on UT 2015 December 19 at the IAO using the 105-cm Murikabushi Cassegrain telescope. The observatory is located on Ishigaki Island, Okinawa Prefecture, Japan (24\arcdeg 22\arcmin 22\arcsec N, 124\arcdeg 08\arcmin 21\arcsec E, 197 m). We made use of the dataset obtained from the observation, comparing it with the the same-day OAO observations, focusing on the comet's morphology and radial profile. For this observation, we exploited another set of MITSuME, which was designed identically to that for the OAO. It covers a FOV of 12\arcmin $\times$ 12\arcmin\ with a pixel resolution of 0.72\arcsec. The average FWHM was $\sim$ 4\arcsec, which is slightly larger than that of the OAO on the same day. In a similar manner to the OAO, we employed the comet tracking mode of the IAO.

\subsection{Data Reduction and Analysis}

We obtained two days of polarimetric data and one day of spectroscopic data from the HHO observations, three days of imaging data from the OAO observations, and one day of imaging data from the IAO observations (see Table 1). The raw observational data from the OAO and IAO were reduced by means of standard pre-processing methods: bias or dark subtractions and flat fielding in IRAF. For the HHO data, we exploited the HONIR data reduction pipeline for trimming of the over-scanned regions, bias and dark subtractions, flat fielding, cosmic ray removal, and bad pixel corrections.

To derive the polarization degrees using the HONIR data, we first subtracted the background sky intensity from each image. Since the cometary tail extended in southeast and northwest directions (see Figure 1), we derived the average count value at a distant place in the northwest direction (approximately 2.5 $\times$ $10^5$ km from the nucleus) and regard it as the background sky intensity. The HONIR optics split incident light into two different windows seen from extraordinary and ordinary lights and enable us to obtain a chain of ten columns in a single image (see Figure 11 in \citealt{Akitaya et al.2014}). Thus, we derived $P$ by performing photometry of two focused images. Since the comet was sufficiently bright (approximately 7.0 mag in $R_{\rm C}$, with S/N $\sim$ 1100, and approximately 5.9 mag in $K_{\rm S}$, with S/N $\sim$ 175, within a 10\arcsec\ circular aperture with a single exposure), we derived $P$ from one set of data, which consists of four images taken at the HWP position angles of 0\arcdeg, 45\arcdeg, 22.5\arcdeg, and 67.5\arcdeg\ in a row, and computed nightly averaged $P$ values. 

A derivation of $P$ requires additional corrections associated with instrumental effects. By using data of a strongly polarized star, BD+59d389, and an unpolarized star, HD 212311 \citep{Schmidt et al.1992}, taken on UT 2015 December 17, we examined the instrumental polarization, polarization efficiency, reference position angle of the polarization vector, and corresponding errors; we found that these values are consistent with those of \citet{Akitaya et al.2014}. We corrected the polarization efficiencies at each filter and the instrumental polarization ($\lesssim$ 0.2 \%) and revised the position angle offsets $\Delta$$\theta$ between the fiducial axis of HONIR and the fast axis of polarized emission (i.e., polarization angle) from the observed values. We derived the ratio of the Stokes parameters normalized by the intensity, $Q/I$ and $U/I$, and propagated errors in the same manner as \citet{Kawabata et al.1999} and \citet{Kuroda et al.2015}.

The derived $P$ and the position angle of the strongest electric vector $\theta_{\rm P}$ were converted into quantities relative to the scattering plane (the plane on which the Sun, Earth, and comet are located) as follows: 

\begin{eqnarray}
P_{\rm r} = P\cos{(2\theta_{\rm r})},
\label{eq:eq1}
\end{eqnarray}
\noindent and
\begin{eqnarray}
\theta_{\rm r} = \theta_{\rm P} - (\phi\pm90\arcdeg),
\label{eq:eq2}
\end{eqnarray}
\noindent where $\phi$ is the position angle of the scattering plane, whose sign is manipulated to satisfy $0\arcdeg$ $\le$ ($\phi$ $\pm$ 90\arcdeg) $\le$ $180\arcdeg$ \citep{Chernova et al.1993}. From Eqs. (\ref{eq:eq1})--(\ref{eq:eq2}), we obtained $P_{\rm r}$=10.2 $\pm$ 1.6 \% in the $R_{\rm C}$ band, 12.2 $\pm$ 1.0 \% in the $I_{\rm C}$ band, and 14.9 $\pm$ 0.8 \% in the $K_{\rm S}$ band on UT 2015 December 17 and $P_{\rm r}$=11.5 $\pm$ 1.7 in $R_{\rm C}$ and 15.1 $\pm$ 1.0 in $K_{\rm S}$ on December 18. We tabulated the resulting polarimetric parameters in Table 2. Note that these polarimetric values are provisional and less creditable for the study of dust polarimetry because the observed signals are mixtures of dust and gas components, as shown in Section \ref{sec:sec31}.

\begin{deluxetable}{ccccccccc}[h!]
\tablecaption{Nightly averaged polarimetric results \label{information}}
\tablehead{
\multirow{2}{*}{Date} & \multirow{2}{*}{Filter} & $P$[\%] & $\sigma_{\rm P}[\%]$ & $\theta_{\rm P}$[\arcdeg] & $\sigma_{\theta_{\rm P}}$[\arcdeg] & $P_{\rm r}$[\%] &$\sigma_{P_{\rm r}}$[\%] & $\theta_{\rm r}$[\arcdeg]\\ 
 & & (1) & (2) & (3) & (4)& (5) & (6) & (7)
}
\startdata 
  \multirow{3}{*}{2015/12/17} & $R_{\rm C}$ & 10.41 & 0.41 & 23.55 & 0.70 & 10.18 & 1.59 & -6.05\\ 
  & $I_{\rm C}$ & 12.20 & 0.48 & 31.58 & 1.02 & 12.17 & 0.98 & 1.98\\ 
  & $K_{\rm S}$& 14.86 & 0.72 & 29.29 & 1.98 & 14.86 & 0.79 & -0.31\\
\\
  \multirow{2}{*}{2015/12/18}  & $R_{\rm C}$ & 11.70 & 0.58 & 24.96 & 0.82 & 11.53 & 1.72 & -4.84\\  
  & $K_{\rm S}$ & 15.14 & 0.81 & 29.37 & 2.16 & 15.14 & 0.95 & -0.43\\
\enddata
\tablenotetext{(1)}{ Observed $P$ of the comet}
\tablenotetext{(2)}{ Standard deviation of $P$}
\tablenotetext{(3)}{ Electric vector position angle}
\tablenotetext{(4)}{ Standard deviation of $\theta_{\rm P}$}
\tablenotetext{(5)}{ $P$ relative to the scattering plane (see Eq. \ref{eq:eq1})}
\tablenotetext{(6)}{ Standard deviation of $P_{\rm r}$}
\tablenotetext{(7)}{ Position angle relative to the scattering plane (see Eq. \ref{eq:eq2})}

\end{deluxetable}

The spectroscopic data were analyzed with \textbf{\tt apall} in IRAF. Since the signal-to-noise ratio of the NIR spectrum is too low (S/N $<$ 3) to perform scientific analyses, we only analyzed the optical spectrum. We extracted the spectrum of the comet in a rectangular area within a box of length 9500 km (11.0\arcsec) along the N-S direction and 1900 km (2.2\arcsec) along the E-W direction. We chose the large aperture along the slit because the comet moved by 4.8\arcsec\ during a single exposure along the N-S direction due to the unavailability of the comet-tracking mode. The sky was taken to be the average value measured both from the north and south $1\times10^4$ km symmetrically away from the center. For flux calibration, we observed the spectroscopic standard star HR5501 (B9.5V) \citep{Hamuy et al.1992, Hamuy et al.1994}. We extracted the spectral intensity of the object within an effective area of 24.2  arcsec$^2$ (i.e., 2.2\arcsec$\times$4.8\arcsec) centered on the photocenter of the comet (i.e., the position of the nucleus). 

For the analyses of imaging data obtained from OAO and IAO observations, we converted the pixel coordinates into celestial ones using WCSTools \citep{Mink1997} and matched the locations of the stellar objects with those of USNO-A2 catalog stars \citep{Monet1998}. These observed raw data were pre-processed in a standard manner with IRAF (bias and dark subtraction and flat fielding). To perform aperture photometry, we utilized the APPHOT package in IRAF, calibrating target magnitudes with UCAC-3 catalog field stars \citep{Zacharias et al.2009}. We assumed the systematic error of 0.1 mag for calibration. Finally, to improve the S/N of these images and remove background stars, we median-combined the data using the instantaneous location of the comet as the center.

\section{Results}

\begin{figure}[ht!]
 \epsscale{1.0}
  \plotone{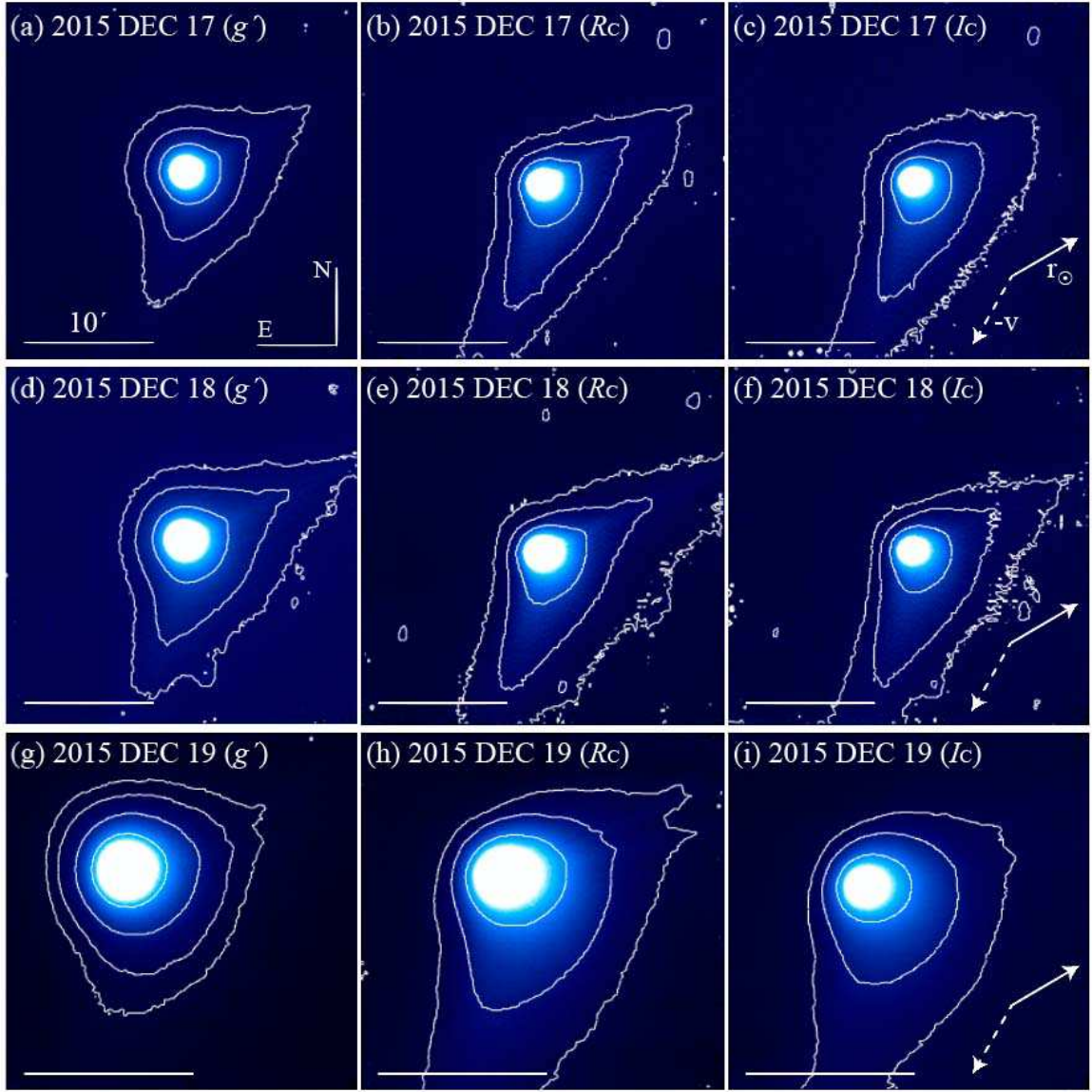}
  \caption{Median-combined images of C/2013 US10 in the {\sl g}$'$, $R_{\rm C}$, and $I_{\rm C}$ bands. The images are all aligned in the standard orientation, i.e., north is up and east is to the left. We superimposed white bars of scale 10\arcmin\ and contours of intensity with a logarithmic scale on each image. The white solid and dashed arrows denote the Sun-target radius vectors and the negative heliocentric velocity vectors of the target, respectively. (a) through (c) were taken on UT 2015 December 17 at the OAO, (d) through (f) on December 18 at the OAO, and (g) through (i) on December 19 at the IAO. The first, second, and third columns show the images taken in the {\sl g}$'$, $R_{\rm C}$, and $I_{\rm C}$ filters, respectively.}\label{fig:composite} 
\end{figure}

Figure \ref{fig:composite} shows the multi-band images of C/2013 US10. These images exhibit a spherical coma near the nucleus, in addition to ion and dust tails. The ion tail extends along the northwest direction (approximately in the anti-solar direction, $\bf{r_\odot}$), whereas the prominent dust tail extends southeast (approximately along the negative velocity vector, $\bf{-v}$). In this section, we estimate the intensity ratio of the gas and dust components (Section \ref{sec:sec31}), and derive the polarization degrees that originate solely from the dust particles in the coma. Moreover, we examine the dependencies of $P_{\rm r}$ on the cometocentric distance (Section \ref{sec:aperture}), phase angle (Section \ref{sec:sec33}), and wavelength (Section \ref{sec:sec34}).

\subsection{Estimation of the Influence of Gas in Standard Optical Filters}
\label{sec:sec31}
\begin{figure}[ht!]
 \epsscale{1.0}
  \plotone{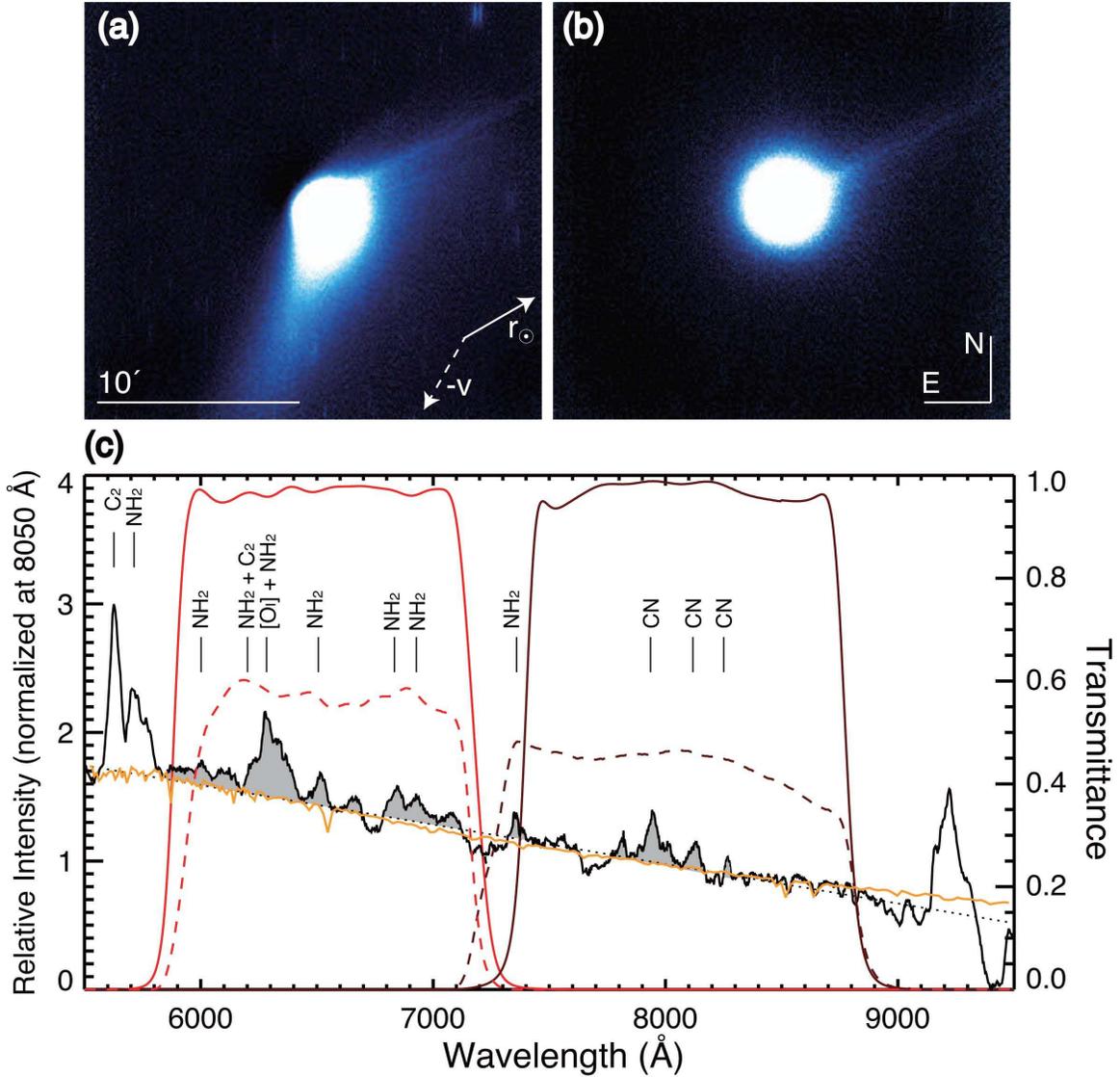}
  \caption{Dust (a) and gas (b) intensity maps in the $R_{\rm C}$ band with the symbols and scales illustrated similarly to Figure \ref{fig:composite}.~(c) presents a median-combined spectrum of C/2013 US10 from 5500 \AA\ to 9500 \AA\ normalized at 8050 \AA. We identified the discernible gas emission lines and colored the emission regions that we utilized to derive the gas contamination rates in the $R_{\rm C}$ and $I_{\rm C}$ filters gray. The black dotted line denotes the continuum fit, the orange curve denotes the solar spectrum at zero airmass, and the red and brown solid and dashed curves illustrate the efficiencies of the HONIR and MITSuME filters, respectively. The spike near 9200--9300 \AA\ is an artifact of HONIR.}\label{fig:photspec} 
\end{figure}

To eliminate the effect of gas emission in the observed data, we took advantage of the multi-band images in conjunction with the spectroscopic data by cross-checking the two results, which were independently obtained on the same epoch. First, for the estimation of gas influence in the coma, we followed a technique from \citet{Ishiguro et al.2016} (Section 3.5 therein). Using the fact that the contribution of gas emission is more significant in {\sl g}$'$ than in the $R_{\rm C}$ and $I_{\rm C}$ bands \citep{Meech2004}, we subtracted the dust component (which was initially produced using $I_{\rm C}$ and {\sl g}$'$ images) from the $R_{\rm C}$-band image. Adjusting the sky level equivalently, we multiplied the dust map image by a conversion factor to eliminate the dust tail from the differential image. We regarded this differential image as an intensity map of the gas emission components. Conversely, we derived the dust continuum map in the $R_{\rm C}$ band by subtracting the above gas intensity map from the composite $R_{\rm C}$-band image. We iterated this process to purify the dust and gas signals separately. Figures \ref{fig:photspec}-(a) and -(b) show the resultant maps of the isolated dust and gas components. Although we distinguished between dust and gas signals, we found that the dust intensity map exhibits not only the dust continuum component but also a faint ion tail (Figure \ref{fig:photspec}-(a)). Similarly, the gas intensity map exhibits an ion tail in addition to a spherical neutral gas emission component (Figure \ref{fig:photspec}-(b)). Because the ion tail is faint ($\sim$11 \% of the dust signal in the northeast tail but negligible in the inner coma and southwest tail), we assume that the ion tail has no significant influence in the following derivation of the dust polarization degree.

For the sake of inter-comparison of the results obtained from imaging and spectroscopic data, we used an identical effective aperture of 24.2 arcsec$^2$ (i.e., 11\arcsec\ along N-S and 2.2\arcsec\ along E-W) for the above imaging analysis. To derive the gas emission rate, $f^{{\rm g}}_{\lambda}$, over the total intensity of the image, we used the following equation:
\begin{eqnarray}
f^{{\rm g}}_{\lambda} \equiv \frac{I^{\rm g}_{\lambda}}{I^{\rm g+d}_{\lambda}}=\frac{I^{\rm g}_{\lambda}}{I^{\rm g}_{\lambda}+I^{\rm d}_{\lambda}}~,
\label{eq:eq3}
\end{eqnarray}
\noindent where the subscript $\lambda$ is the wavelength (either the $R_{\rm C}$ or $I_{\rm C}$ bands in our study). $I^{\rm g}_{\lambda}$ and $I^{\rm d}_{\lambda}$ denote the fluxes of the gas and dust components, and $I^{\rm g+d}_{\lambda}$ indicates the sum of these two at $\lambda$. From our observational data (i.e., Figure \ref{fig:photspec} (a)--(b)), we obtained $f^{{\rm g}}_{R_{\rm C}}$ = 8.8 $\pm$ 1.0 \% and $f^{{\rm g}}_{I_{\rm C}}$ = 3.2 $\pm$ 0.4 \%.

Figure \ref{fig:photspec}-(c) shows the comet spectrum on UT 2015 December 18. We plotted the relative intensity from 5500 {\AA} to 9500 {\AA} normalized at 8050 \AA\ with the extraterrestrial solar spectrum \citep{Gueymard2004}. We labeled prominent gas emission lines and colored the regions where we detected excess flux relative to the continuum gray. To calculate the fraction of gas flux in the $R_{\rm C}$ and $I_{\rm C}$-filter domains from the spectrum, we used the following integral equations:
\begin{eqnarray}
f^{{\rm g}}_{R_{\rm C}} = \frac{\int_{\lambda_{\rm 1}}^{\lambda_{\rm 2}} \left[I_{obs}(\lambda)-I_{cont}(\lambda)\right]~d\lambda}{\int_{\lambda_{\rm 1}}^{\lambda_{\rm 2}}I_{obs}(\lambda)~d\lambda},
\label{eq:eq4}
\end{eqnarray}
\noindent and
\begin{eqnarray}
f^{{\rm g}}_{I_{\rm C}}= \frac{\int_{\lambda_{\rm 3}}^{\lambda_{\rm 4}} \left[I_{obs}(\lambda)-I_{cont}(\lambda)\right]~d\lambda}{\int_{\lambda_{\rm 3}}^{\lambda_{\rm 4}}I_{obs}(\lambda)~d\lambda},
\label{eq:eq5}
\end{eqnarray}
\noindent where $I_{obs}(\lambda)$ and $I_{cont}(\lambda)$ denote the observed flux and continuum flux (dotted line in Figure \ref{fig:photspec}-(c)). We set $\lambda_{\rm 1}$=5800 {\AA} and  $\lambda_{\rm 2}$=7200 {\AA} for the $R_{\rm C}$ domain and $\lambda_{\rm 3}$=7400 {\AA} and $\lambda_{\rm 4}$=8800 {\AA} for the $I_{\rm C}$ domain. We obtained $f^{{\rm g}}_{R_{\rm C}}$ = 8.3 $\pm$ 0.5 \% and $f^{{\rm g}}_{I_{\rm C}}$ = 3.0 $\pm$ 0.1 \%. The gas fractions in the $R_{\rm C}$ and $I_{\rm C}$ bands are in agreement with those obtained from the imaging approach ($f^{{\rm g}}_{R_{\rm C}}$ = 8.8$\pm$1.0 \% and $f^{{\rm g}}_{I_{\rm C}}$ = 3.2 $\pm$ 0.4\%) to the accuracy of our measurements. Although we only have spectroscopic data taken on UT 2015 December 18, we assumed that the gas production rates did not change abruptly within a day because there is no significant variation between the images taken on December 17 and 18. Accordingly, we adopted the same gas emission ratios to correct the $P_{\rm r}$ value on December 17 in the following sections.

\subsection{Cometocentric Distance Dependence of $P_{\rm r}$}
\label{sec:aperture}
\begin{figure}[ht!]
 \epsscale{1.0}
  \plotone{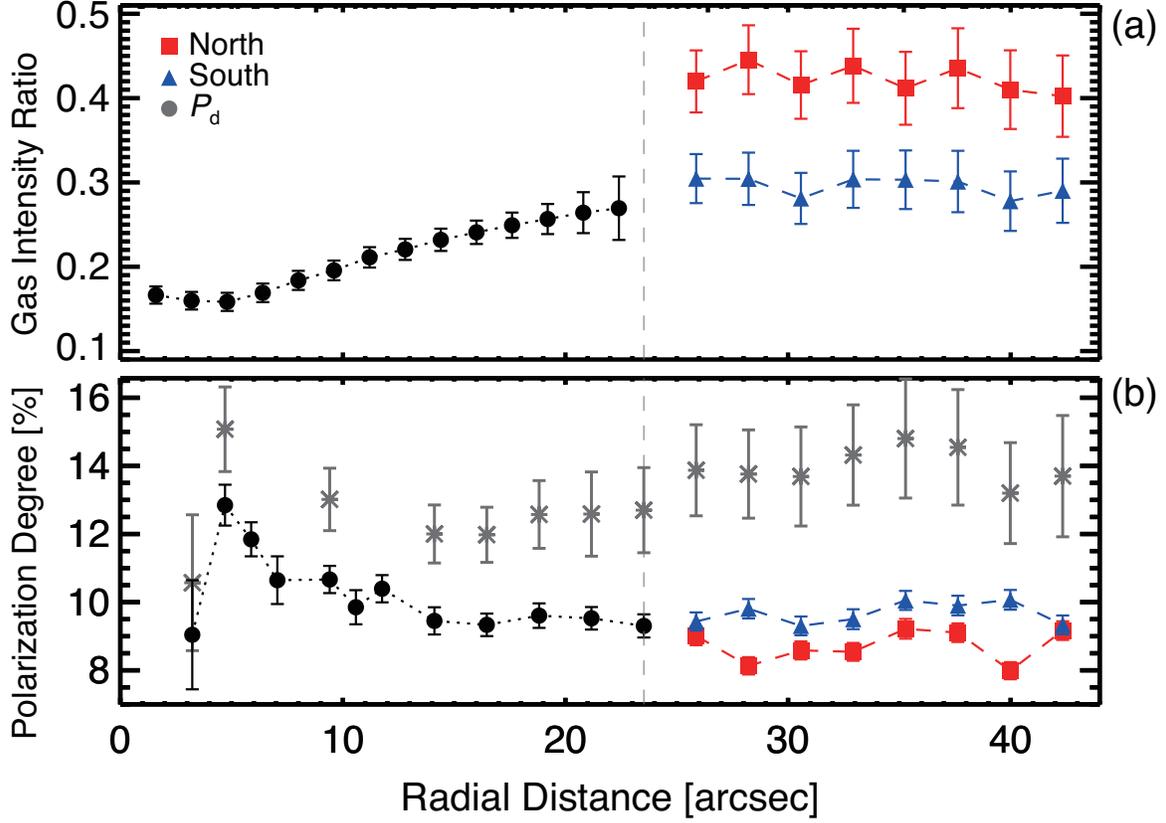}
  \caption{Radial profiles of (a) the gas intensity ratio, $f^{{\rm g}}_{R_{\rm C}}$, and (b) $P_{\rm r}$ as a function of the apparent distance from the nuclear position. The vertical dashed lines at 23\arcsec\ denote the boundary of the different data analyses (see the text). Within the boundary, we conducted circular aperture polarimetry. Beyond this boundary, we divided the sky regions into two areas (north and south wings), treating them separately to derive these two parameters. The gray asterisks in (b) denote $P_{\rm r}^{\rm d}$, which are corrected with the corresponding values of (a).}\label{fig:radpolgas} 
\end{figure}

We examined the cometocentric distance dependence of $P_{\rm r}$. To secure the best-S/N data possible for this analysis, we took advantage of the $R_{\rm C}$-band data on UT 2015 December 17. Within the area defined by a cometocentric distance $<$ 23\arcsec\ ($2\times10^4$ km), we performed differential aperture polarimetry with a cadence of 5 pixels (1.5\arcsec) to derive $P_{\rm r}$. At the cometocentric distance $\gtrsim$ 23\arcsec, we derived the averaged $P_{\rm r}$ within an 8 pixel-by-8 pixel square (2.4\arcsec$\times$2.4\arcsec) region from the nucleus to the given position along the N-S direction (aligned with the long axis of the polarization mask of HONIR) because of the limited column width of the HONIR FOV (47\arcsec). We derived the values in the south and north regions. At the same time, we derived the gas intensity ratio ($f^{{\rm g}}_{R_{\rm C}}$) to examine the correlation with $P_{\rm r}$ in the same manner as we derived it from Eq. (\ref{eq:eq3}). The results are shown in Figure \ref{fig:radpolgas}.

In Figure \ref{fig:radpolgas}, we show the profiles of $f^{{\rm g}}_{R_{\rm C}}$ (a) and $P_{\rm r}$ (b) with respect to the apparent distance from the nuclear position. $P_{\rm r}$ of the comet varies from 8 \% to 13 \%. This variation of the observed polarization degree straddles the gas- and dust-rich groups at the given phase angle (shown later in Figure \ref{fig:polred}). The gas intensity ratio gradually increases within 23\arcsec\ (corresponding to $\sim$~2$\times$$10^4$ km) but remains almost constant beyond 23\arcsec\ for both the north and south regions. There is a vertical offset in $f^{{\rm g}}_{R_{\rm C}}$ and $P_{\rm r}$ between the north and south components beyond 23\arcsec. Because dust particles were accelerated toward the south--east direction via radiation pressure, the $f^{{\rm g}}_{R_{\rm C}}$ values in the south are significantly less than those in the north. Conversely, $P_{\rm r}$ exhibits the opposite trend in the north and south regions. This evidence implies that there is a correlation between $f^{{\rm g}}_{R_{\rm C}}$ and $P_{\rm r}$ within this region.

\begin{figure}[ht!]
 \epsscale{1.0}
  \plotone{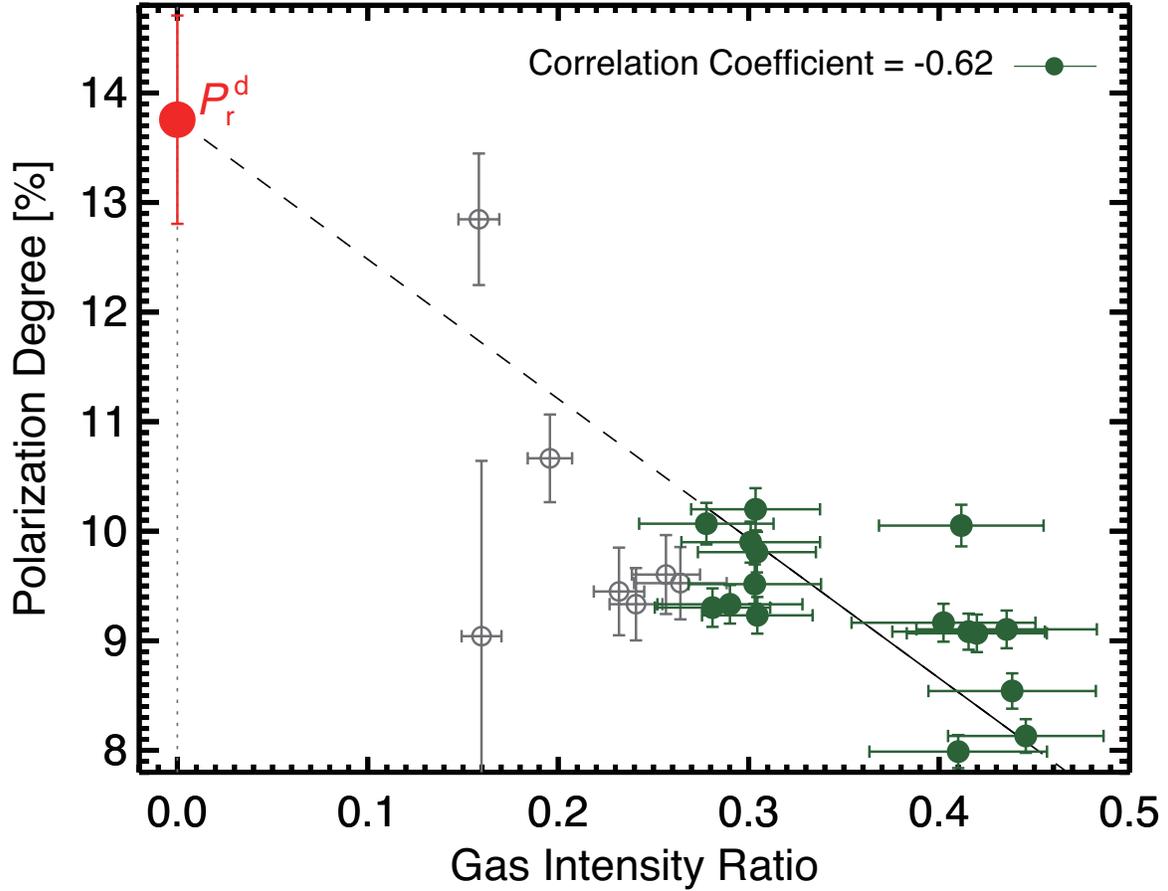}
  \caption{Correlation between the gas intensity ratio $f^{{\rm g}}_{R_{\rm C}}$ and $P_{\rm r}$ of the comet in the $R_{\rm C}$ band. We utilized the data obtained separately from the northern and southern parts of the center of the comet beyond the region of 23\arcsec\ from the nucleus. The y-intercept (red circle) represents the estimated $P_{\rm r}^{\rm d}$ of the gas-free (i.e., dust-only) component in the cometary coma extrapolated from the observed data (green circles). The open gray circles are the data from the inner region of 23\arcsec\, which we did not include in the fitting.}\label{fig:anticorr} 
\end{figure}

Figure \ref{fig:anticorr} presents the correlation between $f^{{\rm g}}_{R_{\rm C}}$ and $P_{\rm r}$. The plot indicates a reliable correlation with a coefficient of $-$0.62 beyond 23\arcsec\ (filled circles) and scatter within 23\arcsec\ (open circles). We utilized the data beyond 23\arcsec\ and employed the FITEXY estimator for linear fitting considering both x- and y-axis errors \citep{Press et al.1992,Tremaine et al.2002}. That is, we can say that the gas contamination is dominant such that the derived $P_{\rm r}$ varies depending on the aperture size in this region, as suggested in previous research \citep{Chernova et al.1993,Jockers1997,Jockers et al.2005}. From the fitting, we obtained a gas-free polarization degree of $P_{\rm r, R_{\rm C}}^{\rm d}$ = 13.8 $\pm$ 1.0 \% (the y-intercept in Figure \ref{fig:anticorr}) and a gas polarization degree of $P_{\rm r, R_{\rm C}}^{\rm g}$ = 1.0 $\pm$ 2.7 \% ($P_{\rm r}$ at $f^{{\rm g}}_{R_{\rm C}}$=1) via extrapolation of the fitting line. In a similar manner, we obtained $P_{\rm r, I_{\rm C}}^{\rm d}$ =12.5$\pm$1.1 \% and $P_{\rm r, I_{\rm C}}^{\rm g}$ = $-$3.1$\pm$3.8 \%. The large uncertainties of the $P_{\rm r}^{\rm g}$ that are also responsible for an unrealistic negative value of $P_{\rm r}^{\rm g}$ result from the limited gas intensity ratio coverage of our data. In addition, we consider that the errors of gas polarization become larger due to the unavailability of narrow-band data (that is, different types of gas components could be blended in these broadband data). In fact, the Pearson correlation coefficient is -0.62, which is moderately significant but not ideal. Despite the uncertainties, as we will show in Section 3.3, our results of dust polarization degrees are consistent with the ones of the previous studies at the given phase angle, which supports our simple manipulation of the wide multi-band data as one of the tools for measurement of $P_{\rm r}^{\rm d}$.

Using the measured gas polarization degree of $P_{\rm r,\lambda}^{\rm g}$, we derived the `gas-free polarization degree of dust particles' at an arbitrary location using the following equation:
\begin{eqnarray}
P_{\rm r,\lambda}^{\rm obs} = P_{\rm r,\lambda}^{\rm d}~f_\lambda^{\rm d} + P_{\rm r,\lambda}^{\rm g}~f_\lambda^{\rm g},
\label{eq:eq6}
\end{eqnarray}
\noindent where $P_{\rm r,\lambda}^{\rm obs}$,  $P_{\rm r,\lambda}^{\rm d}$, and $P_{\rm r,\lambda}^{\rm g}$ denote the $P_{\rm r}$ of the dust plus gas (the observed quantity), dust-only, and gas-only components, respectively, at a given wavelength $\lambda$. The results are shown in Figure \ref{fig:radpolgas}-(b) (asterisks). The cometocentric dependence of $P_{\rm r}^{\rm d}$ after gas correction becomes $\leq$3 \%, exhibiting a shallow bump at 4.7\arcsec\ (corresponding to $\sim$4,100 km from the center). Although the small number of pixels restricts the interpretation of the inner part of the bump, the overall behavior of $P_{\rm r}^{\rm d}$ is almost constant over the wide area within a cometocentric distance of 5 $\times$ $10^{4}$ km.

\subsection{Phase Angle Dependence of $P_{\rm r}$}
\label{sec:sec33}
To determine the polarimetric class of C/2013 US10, we plot $P_{\rm r}$ integrated within an aperture radius of 3500 km from the comet center with respect to $\alpha$ together with the values for other comets. Note that this polarization degree is averaged one which does not include the information of spatial variations of $P_{\rm r}$ within the aperture size. We selected $P_{\rm r}$ of other comets that were based on narrow-band filters to minimize possible gas influences (the comet names and references are provided in the captions). We grouped the wavelength domains as follows: the red domain, when data were taken at a wavelength in the range of 6200 {\AA} $<$ $\lambda$ $<$ 7300 {\AA}; the $I$-filter domain, 7300 {\AA} $<$ $\lambda$ $<$ 9000 {\AA}; and the $K$-filter domain, 20000 {\AA} $<$ $\lambda$ $<$ 25000 {\AA}. To facilitate a comparison between ours and the previous observations, we employed an empirical trigonometric function to fit the data \citep{Penttila et al.2005}:
\begin{eqnarray}
P_{\rm r}(\alpha) = b (\sin \alpha)^{c_{\rm 1}}\times\sin(\alpha - \alpha_{\rm 0})\times \cos \Big(\frac{\alpha}{2}\Big)^{c_{\rm 2}}~,
\label{eq:eq7}
\end{eqnarray}
\noindent where $b$, $c_{\rm 1}$, $c_{\rm 2}$, and $\alpha_{\rm 0}$ are wavelength-dependent parameters to characterize the phase angle dependence. The following parameters were obtained through fitting the data: $b$ $=$ 33.25 \%, $c_{\rm 1}$ $=$ 0.85, $c_{\rm 2}$ = 0.39, and $\alpha_{\rm 0}$ $=$ 21.90\arcdeg\ for the high-$P_{\rm max}$ class and $b$ $=$ 17.78 \%, $c_{\rm 1}$ $=$ 0.61,  $c_{\rm 2}$ $=$ 0.14, and $\alpha_{\rm 0}$ $=$ 21.90\arcdeg\ for the low-$P_{\rm max}$ class in the $R_{\rm C}$ band and $b$ $=$ 39.21 \%, $c_{\rm 1}$ $=$ 0.69, $c_{\rm 2}$ $=$ 0.65, and $\alpha_{\rm 0}$ $=$ 23.41\arcdeg\ in the $I_{\rm C}$ band. Next, we applied $P_{\rm r, R_{\rm C}}^{\rm g}$ = 1.0 $\pm$ 2.7 \% and $P_{\rm r, I_{\rm C}}^{\rm g}$ = $-$3.1$\pm$3.8 \% (see Section \ref{sec:aperture}) for gas correction to the averaged $P_{\rm {r}}$ values and obtained $P_{\rm {r},\it R_{\rm C}}^{\rm d}$=11.1$\pm$1.9 \% in the $R_{\rm C}$ band and $P_{\rm {r},\it I_{\rm C}}^{\rm d}$=12.5$\pm$1.1 \% in the $I_{\rm C}$ band on UT 2015 December 17, in addition to $P_{\rm {r},\it R_{\rm C}}^{\rm d}$=12.5$\pm$1.9 \% in the $R_{\rm C}$ band on UT 2015 December 18. For quick comparison with the $P_{\rm {r}}^{\rm d}$ of the previous studies, we adopted $P_{\rm r}^{\rm g}$ = 2.5 \% for NH$_{\rm 2}$ \citep{Kiselev et al.2004} and 4.6 \% for C$_{\rm 2}$ \citep{Ohman1941} molecules calculated at $\alpha$ = 52.1\arcdeg. As a result, we obtained $P_{\rm r}^{\rm d}$ = 10.9$\pm$1.9 \% and $P_{\rm r}^{\rm d}$ = 12.4$\pm$1.1 \% in $R_{\rm C}$ and $I_{\rm C}$ bands, respectively, showing a general agreement of ours and previous ones. This consistency, again, supports our methodology as one of the tools for measurements of the dust polarization degree of comets.

\begin{figure}[ht!]
 \epsscale{1.0}
  \plotone{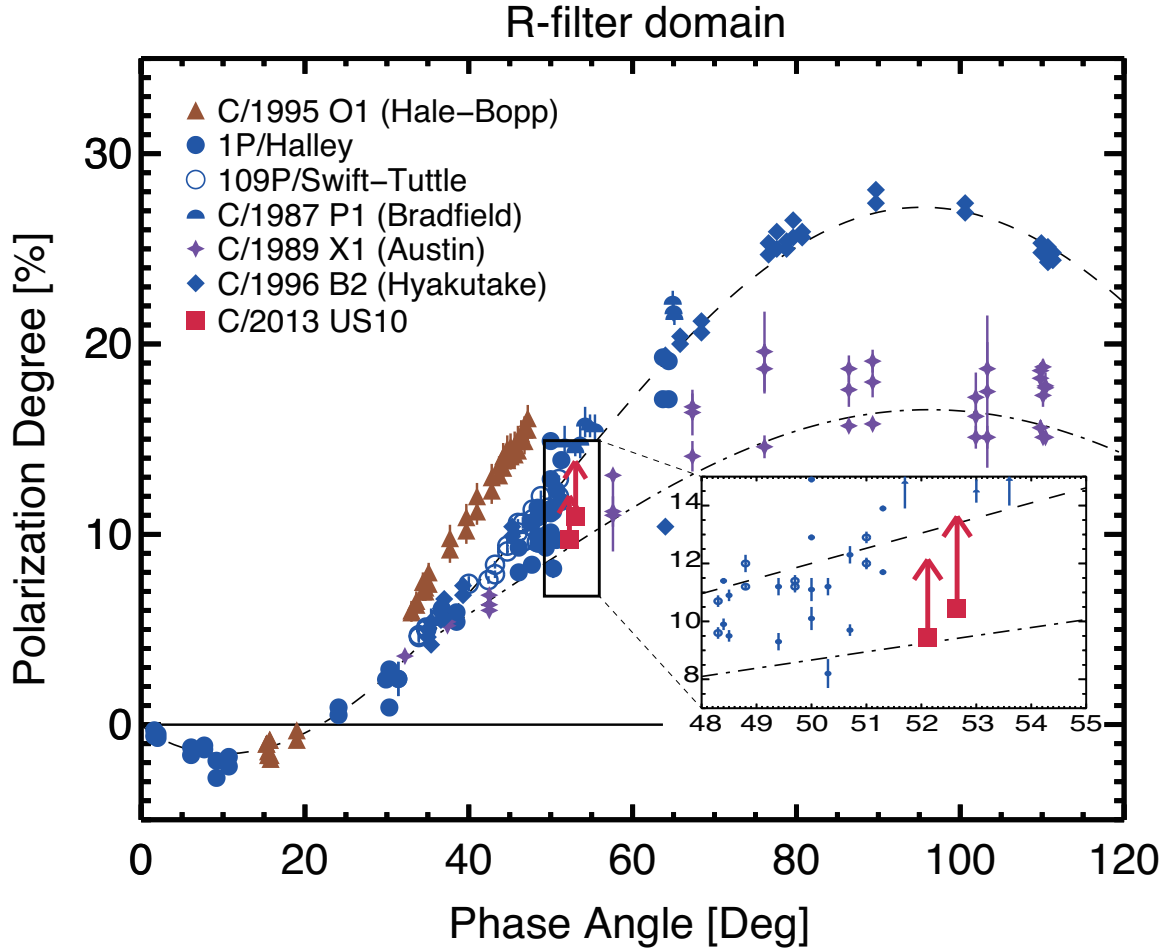}
  \caption{Phase angle dependence of $P_{\rm r}$ in the $R$-filter domain (6200 {\AA}~$<$~$\lambda$~$<$~7300 {\AA}). The filled squares denote the observed $P_{\rm r}$ of C/2013 US10 before gas correction. The arrows denote the amount of change of $P_{\rm r}$ after gas subtraction using the derived value of $P_{\rm r}^{\rm g}$ (see Section 2). For comparison, we show the results for 1P/Halley \citep{Kikuchi et al.1987}; C/Bradfield 1987 P1 \citep{Kikuchi et al.1989}; and 109P/Swift-Tuttle, C/Austin 1989 X1, C/Hale-Bopp 1995 O1, and C/Hyakutake 1996 B2 \citep{Kikuchi2006}, with the empirical curves for the high-polarization group (dashed line) and low-polarization group (dot-dashed line).}\label{fig:polred} 
\end{figure}

\begin{figure}[ht!]
 \epsscale{1.0}
  \plotone{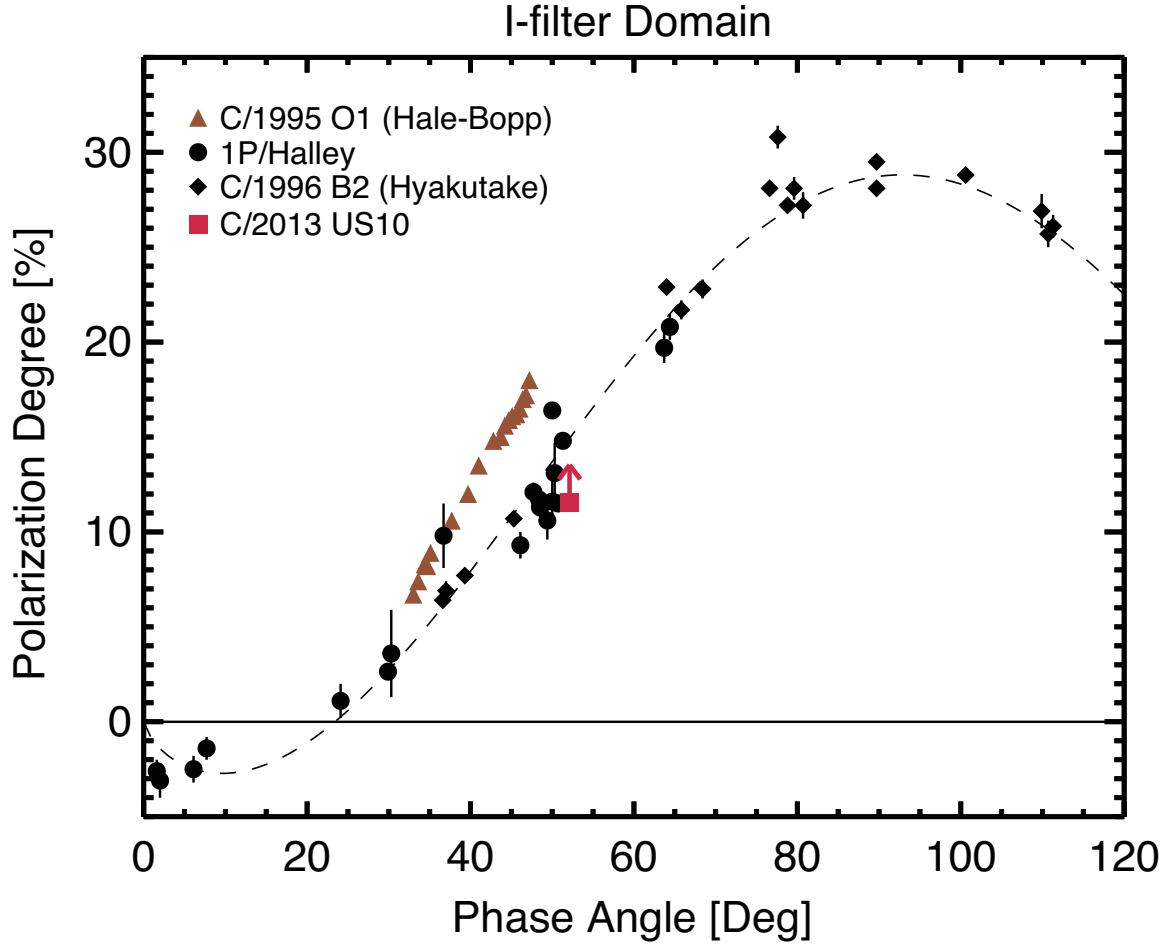}
  \caption{Phase angle dependence of $P_{\rm r}$ in the $I$-filter domain (7300 {\AA}~$<$~$\lambda$~$<$~9000 {\AA}). We plot our results together with archival data for 1P/Halley \citep{Kikuchi et al.1987} and C/1996 B2 (Hyakutake) \citep{Kikuchi2006}.  The notation of the filled squares and arrows is the same as for Figure \ref{fig:polred}. The dashed line denotes the trend line for the high-polarization group comets.}\label{fig:poli} 
\end{figure}

\begin{figure}[ht!]
 \epsscale{1.0}
   \plotone{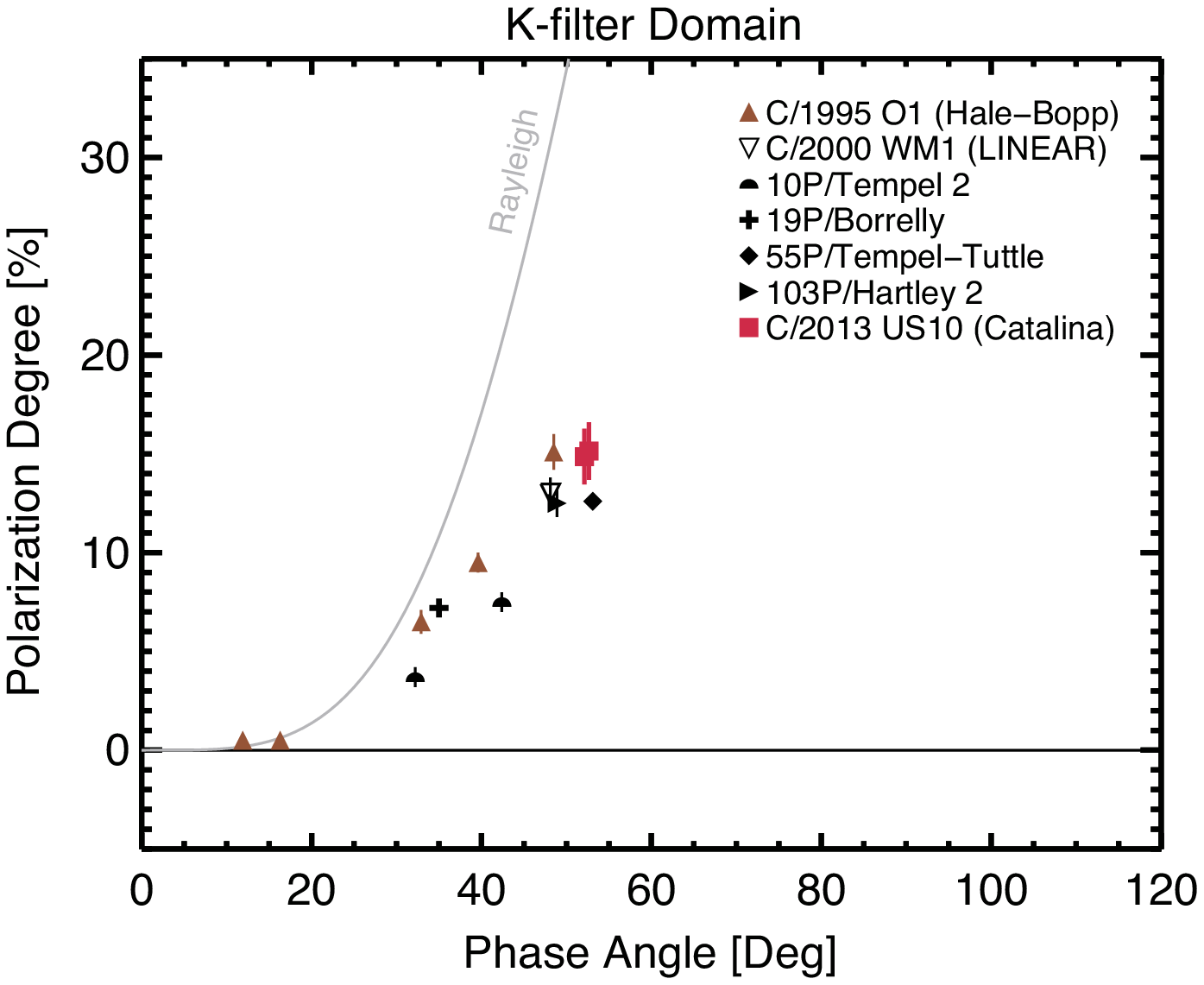}
  \caption{Phase angle dependence of $P_{\rm r}$ in the $K$-filter domain (20000 {\AA}~$<$~$\lambda$~$<$~25000 {\AA}). We plot our observational results as filled red squares over the archival data for 1P/Halley \citep{Kikuchi et al.1987} and C/1996 B2 (Hyakutake) \citep{Kikuchi2006}. We show the line of Rayleigh scattering as a solid curve for reference.}\label{fig:polnir} 
\end{figure}

In the $R$-filter domain (Figure \ref{fig:polred}), our results for C/2013 US10 are located closer to the low-polarization class before the gas correction, whereas they are nearer the high-polarization class after the gas correction. The amount of $P_{\rm r}$ change via gas correction is denoted by the length of the arrow. In the $I$-filter domain (Figure \ref{fig:poli}), our results are consistent with those of so-called dust-rich comets (except the distinctively high C/1995 O1 (Hale-Bopp)). The arrows are plotted here also, but the effect of the gas subtraction is insignificant (within the scatter in the values associated with a single comet) in the red domain. In the $K$-filter domain (Figure \ref{fig:polnir}), all comets, including C/2013 US10, do not fit the line for Rayleigh scattering particles. This suggests that the effective sizes of dust particles in the cometary comae are greater than the observed wavelength. A careful examination of the phase plot in the $K$ band reveals that short-period comets (10P/Tempel 2 and 55P/Tempel-Tuttle) tend to exhibit lower $P_{\rm r}$ values, whereas long-period or non-periodic comets (Hale-Bopp and C/2000 WM1) exhibit higher $P_{\rm r}$. Although there are exceptions of 19P/Borrelly and 103P/Hartley 2 (short-period comets exhibiting high $P_{\rm r}$ in the $K$ band), the moderately high $K_{S}$-band polarization for C/2013 US10 is qualitatively consistent with the high-polarization group of long-period or non-periodic comets in the $K$-filter domain.

\subsection{Spectral Dependence of $P_{\rm r}$}
\label{sec:sec34}
\begin{figure}[ht!]
 \epsscale{1.0}
  \plotone{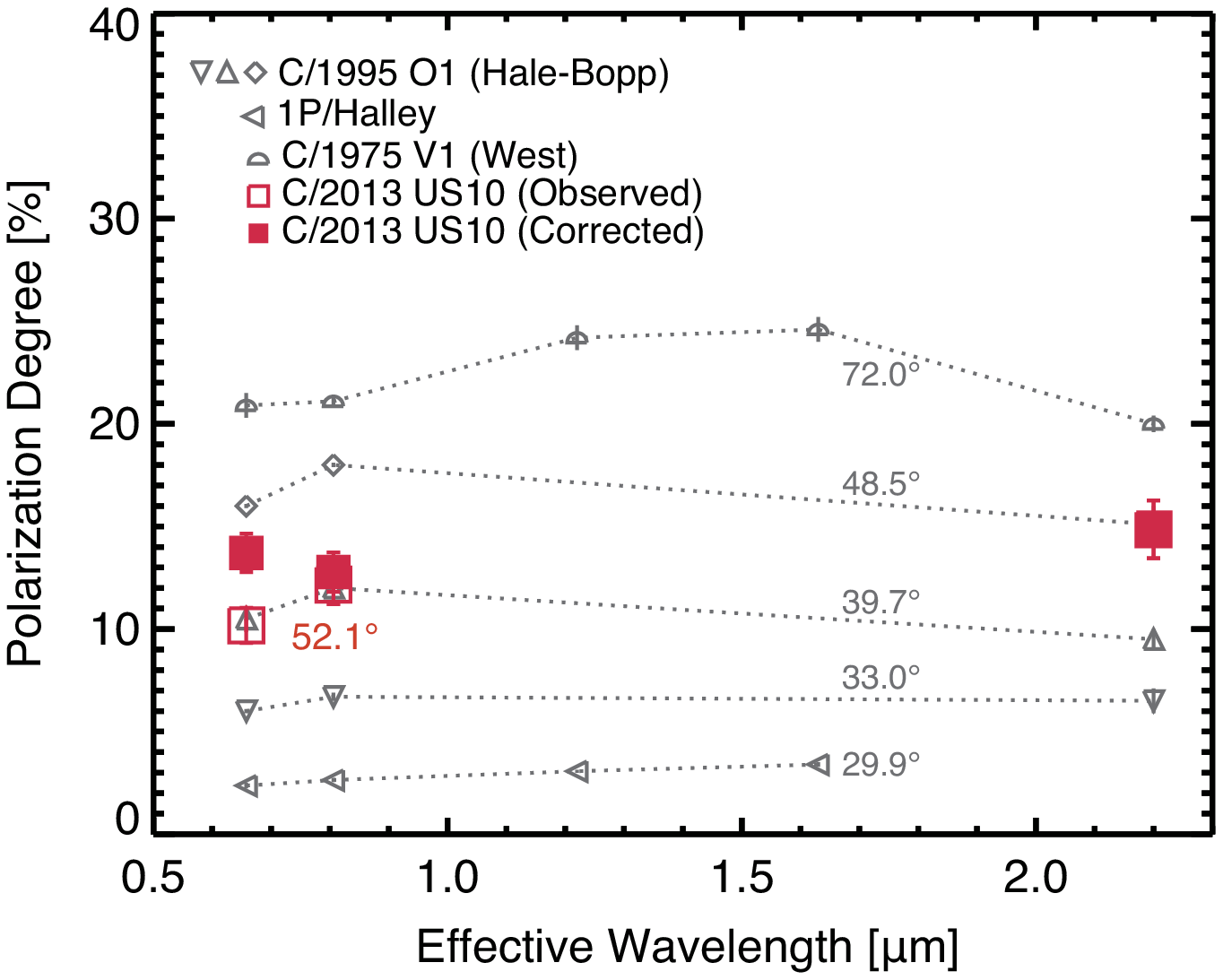}
  \caption{Wavelength dependences of $P_{\rm r}$ of C/2013 US10 compared with C/1995 O1 (Hale-Bopp), 1P/Halley, and C/1975 V1 (West). For this comparative study, we referred to the Database of Comet Polarimetry (DOCP) \citep{Kiselev et al.2006} and references therein. Numerical values next to the lines denote the $\alpha$ values of the data.}\label{fig:polcolor} 
\end{figure}

Figure \ref{fig:polcolor} shows the spectral dependence of $P_{\rm r}$, which is sometimes referred to as the polarimetric color. In the figure, the $P_{\rm r}$ values of our data before and after gas correction (quoted from Section 3.2) are shown as open and filled squares, respectively. For a $K_{\rm S}$-band datum, we use observed $P_{\rm r}$ without any gas correction due to a lack of multi-band data taken in this wavelength and to much less contribution of gas emissions compared than in visual wavelength \citep[e.g.,][]{Jones et al2000}. For comparison, we also plot some data from the Database of Comet Polarimetry (DOCP) \citep[][and references therein]{Kiselev et al.2006}. Numerical values next to the lines denote the phase angles $\alpha$. After correcting for the gas influence, the spectral gradients of $P_{\rm r}$ in the $R_{\rm C}$ and $I_{\rm C}$ bands are shifted from 13.3$\pm$12.5 \% \micron$^{-1}$ to -8.7$\pm$9.9 \%  \micron$^{-1}$. From the $I_{\rm C}$ band through the $K_{S}$ band, it is likely that $P_{\rm r}$ still increases by 1.6$\pm$0.9 \% \micron$^{-1}$, although there is a large wavelength gap in our data between these two bands, which prevents a straightforward interpretation of the polarimetric color of the data. The gas-free optical polarimetric color of C/2013 US10 is bluer (or at least grayer) than the average behavior of other comets in the visible spectrum of $\sim$ 8 \% \micron$^{-1}$ at $\alpha$ $\approx$ 45\arcdeg~\citep{Kiselev et al2004,Kiselev et al.2015}. However, the polarimetric slope between $I_{\rm C}$ and $K_\mathrm{S}$ seems to be steeper than those of comets C/1995 O1 (Hale-Bopp) and C/1975 V1 (West).

\section{Discussion}
\subsection{Dust Tail Polarization}
Polarimetric images provide opportunities for investigating the size-dependent polarization degree. Here, we focus on the polarization of the cometary dust tail because solar radiation pressure results in a size-dependent spatial distribution of dust particles, biasing smaller particles toward the anti-solar direction and larger particles toward the negative heliocentric velocity vector. We consider here a simple model in which dust particles are assumed to be ejected with zero velocity. When dust particles are ejected from a cometary nucleus, they are remapped by the reciprocal effects of solar radiation pressure and solar gravity. This effect is parameterized by $\beta$, which is the ratio of  solar radiation pressure to solar gravity \citep{Burns1979}. It is given by
\begin{eqnarray}
\beta = \frac{KQ_{\rm pr}}{\rho r},
\label{eq:eq8}
\end{eqnarray}
\noindent where $r$ is the radius of a spherical dust grain in meters, $\rho$ is the mass density in ${\rm kg ~m}^{\rm -3}$, and $K$ = 5.7 $\times$ $10^{\rm -4}$~${\rm kg~m}^{\rm -2}$ is a constant in the solar gravitational and radiation field. $Q_{\rm pr}$ denotes a dimensionless coefficient of the radiation pressure. We assumed a compact spherical particle with $\rho=1\times10^3~${\rm kg~m}$^{-3}$ and $Q_{\rm pr}$=1 \citep{Ishiguro2008}. With Eq. (\ref{eq:eq8}), we can specify the locus of dust particles having different sizes (syndyne) and ejection epochs (synchrone) \citep{Finson et al1968}. 

\begin{figure}[ht!]
 \epsscale{1.0}
  \plotone{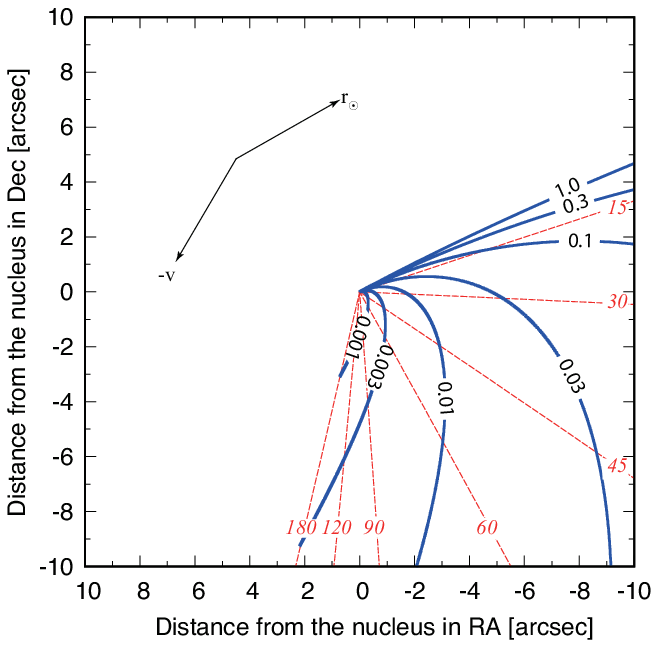}
  \caption{Synchrone and syndyne curves for the C/2013 US10 image on UT 2015 December 17 in comparison with Figure 2-(a). The red dashed lines are the synchrones, and the blue continuous lines are syndynes. The labels on the synchrones denote the days of dust ejection prior to the observation. The syndynes denote $\beta$ values (labeled) increasing anticlockwise from the leftmost $\beta$=0.001 ($r\sim$600 \micron) to $\beta$=1.0 ($r\sim$0.6 \micron). Similar to Figure 2-(a), the antisolar direction ($\bf{r_\odot}$) and the negative heliocentric velocity vector on the celestial plane ($\bf{-v}$) are shown.}\label{fig:syn} 
\end{figure}

Figure \ref{fig:syn} shows the synchrone and syndyne curves. Due to a moderately inclined orbital plane angle (an angle between observer and the cometary orbital plane of 24.8\arcdeg),  these curves are well separated in the observed frame. In our observed image, there are two prominent tails: a brighter one extended to the southeast and a fainter one extended to the northwest. From the comparison between Figure \ref{fig:syn}  and Figure \ref{fig:photspec} (a), it is likely that the primary southeast tail consists of large dust particles (100 \micron\ $<$ $r$ $<$ 1 mm) ejected more than 90 days before the observation. The dust tail becomes fainter with decreasing grain size.  The northwest tail likely consists of small dust particles ($\lesssim$10 \micron) and/or fluffy porous dust particles \citep{Mukai1992}. In addition, ionized gas species overlap the dust tail because the ionized species emit even in the $I_{\rm C}$ band \citep{Sivaraman1973,Yagi2015}. We conjecture that the northwest tail is dominated by ionized gas emission due to the narrowness. 

We examined the polarization degree of the tails. After the consideration of gas contamination and subtraction in the same manner as described in Section \ref{sec:sec31}, we determined the polarization degree of the southeast dust tail to be $P_{\rm {r},\it R_{\rm C}}^{\rm d}$=14.4$\pm$2.6\%. This value is close to the polarization degree in the outer coma (13.8$\pm$1.0 \%), where large and small particles are mixed. From the presence of the northwest dust tail (Figure \ref{fig:syn}) and the gas emission lines (Figure \ref{fig:photspec}), it is certain that the comet was active at the time of observation. However, from the existence of the more obvious southwest dust tail and similarity in $P_{\rm r}$, we conjecture that the activity of the comet became low at the time of our observation and large dust particles occupied a large fraction of the cross section in the coma, obscuring the contrast in $P_{\rm r}$ between the coma and southeast tail.

\subsection{Gas Polarization}

The polarization of the molecular emission was explained by resonance fluorescence \citep{Ohman1941,Le Borgne et al.1987}. After \citet{Ohman1941} proposed the theoretical $\alpha$ dependence of the degree of linear polarization of diatomic molecular bands, whose a maximum value is $p_{\rm max}$ = 7.7 \% at $\alpha=90\arcdeg$, \citet{Le Borgne et al.1987} measured those for comet P/Halley, which showed slight deviation from the theoretical curve, with the polarization angle of molecules almost perpendicular to the scattering plane. \citet{Sen et al.1989} presented the positive polarization of C$_{\rm 2}$ $p$ $\sim$ 0.18~\% at a phase angle $\alpha$ = 12.2\arcdeg, \citet{Kiselev et al.2004} also measured $p$ $\sim$ 3~\% for NH$_{\rm 2}$ at $\alpha$ = $61.6\arcdeg$, and \citet{Kikuchi2006} suggested $p_{\rm max}$ $\sim$ 5~\% for C$_{\rm 2}$ at $\alpha$ = 90\arcdeg. Despite their efforts, there has been no consensus regarding the $\alpha$ dependence of the degree of linear polarization of molecular bands. Therefore, to eliminate signals of gas emission in the observed intensity in standard optical filters, we performed simultaneous multi-wavelength imaging observations of C/2013 US10 in conjunction with the spectroscopic observations.

For the estimation of gas contamination rates from the imaging data, we employ the concept that each filter has different sensitivities to the components of the cometary coma. \citet{Meech2004} showed the broadband (3000--9500 \AA) spectra of comet 8P/Tuttle and substantiated that line emission is much more dominant in the short-wavelength region, indicating that the $I_{\rm C}$-band is the most gas-free region among optical bands. Indeed, Figure \ref{fig:composite} assures that the $I_{\rm C}$-band images capture the strongest and thickest development of dust tails, and the {\sl g}$'$-band images more clearly exhibit spherical gas comae than other filters. 

From Figure \ref{fig:anticorr}, we obtained gas polarization degrees of $P_{\rm r}^{\rm g}$ = 1.0$\pm$2.7 \% via extrapolation of the data obtained from polarimetric and multi-band imaging observations. As we mentioned above, the large errors of the $P_{\rm r}^{\rm g}$ result  from the limited gas intensity ratio coverage and probably from unavailability of a narrow-band filter. Nevertheless, our estimate of $P_{\rm r}^{\rm g}$ at the phase angle $\alpha$ = 52.1\arcdeg\ is slightly less than the theoretical prediction of \citet{Ohman1941} of $\sim$ 4.6 \% but is consistent with the measurement of \citet{Kiselev et al.2004}. Considering that NH$_{\rm 2}$ is the most dominant emission line in the $R_{\rm C}$ filter, the consistency between our result and that of \citet{Kiselev et al.2004} supports our methodology to estimate the gas influence in standard optical filters. 

We note the importance of gas correction in describing the two polarimetric classes in the optical band from Figure \ref{fig:radpolgas}-(b). Before the gas correction, the $P_{\rm r}^{\rm d}$ values vary by $\sim$5 \%, decreasing from the center of the nucleus, which is the typical behavior of gas-rich comets \citep{Kolokolova et al.2007}. However, after the correction using $P_{\rm r}^{\rm g}$, the cometocentric distribution of the gas-free polarization degree exhibits a deviation less than 1$\sigma$ from the average value (asterisks in Figure \ref{fig:radpolgas}-(b)). This small variation of $P_{\rm r}^{\rm d}$ versus cometocentric distance is shown to be well matched with the case of dust-rich comets \citep{Kolokolova et al.2007}, and indeed, the corrected values of $P_{\rm r}^{\rm d}$ at a given phase angle correspond to or are even slightly greater than the nominal value of dust-rich comets. Although we performed this correction for one target here, an identical approach can be employed for other so-called gas-rich comets in the optical band to determine the fractional contamination of their polarization signals caused by depolarizing gas emission lines.

\subsection{Inner Coma Polarization}

After dust particles are ejected from a cometary nucleus, they undergo different evolutionary tracks, depending on their physical properties. Consequently, cometary dust in comae may exhibit inhomogeneous radial distributions. In that sense, imaging polarimetry is of special interest to translate the spatial distribution of $P_{\rm r}$ into the physical properties of cometary dust. Our polarimetric data showed an inhomogeneity in the inner coma, displaying the highest-$P_{\rm r}^{\rm d}$ around 5\arcsec--7\arcsec (corresponding to 5 -- 6 $\times$ 10$^3$ km from the nucleus, Figure \ref{fig:radpolgas}-(b)). Similar structures were found by the flyby Giotto / Optical Probe Experiment (OPE) of comet 1P/Halley \citep{Levasseur-Regourd et al.1999} and other ground-based observations of comet 73P/Schwassmann-Wachmann 3 \citep{Jones et al.2008,Hadamcik et al.2016}. Hereafter, we will address the possible mechanisms underlying this distribution based on our observational results described in Section 3.

\textit{1. Local enhancement of small-sized dust particles.} Peculiar structures induced by local enhancement of small-sized dust particles (e.g., jets, arcs, and spirals) can occlude the background coma \citep{Renard et al.1996,Hadamcik et al.2003b,Hadamcik et al.2009}. This superimposing region could be shown by contrast in both the intensity and dust color maps. First, we checked the surface brightness profiles of isolated gas- and dust-components (Figure \ref{fig:photspec}-(a), (b)) as a function of the cometocentric distance but found no peculiar variation to underpin this idea. Next, we checked the color profile of dust particles obtained with the dust maps of $R_{\rm C}$ and $I_{\rm C}$ band images. The color of the dust becomes slightly reddened but remains almost constant as the distances 2\arcsec -- 25\arcsec (corresponding to 0.3 -- 2.3 $\times$ 10$^4$ km from the nucleus) without any anomaly on the average value of 0.64$\pm$0.10. We found the similar trend in some comets; 67P/Churyumov -- Gerasimenko \citep{Solontoi et al.2012} shows almost constant dust color of 0.25$\pm$0.01 up to the radial distance of 25\arcsec ($\le$ 2.9 $\times$ 10$^4$ km from the nucleus center), and comet 168P/Hergenrother \citep{Betzler et al.2017} shows a slow outward increase of 0.05 over the radial distance of 50\arcsec\ ($\le$ 1.6 $\times$ 10$^4$ km from the nucleus center) with the average value of 0.52$\pm$0.03. For now, it is difficult to draw definite conclusions from our case but is clear that there are no unusual variations. Hence, local enhancement of small-sized dust particles seems to be untenable explanation for this inner distribution.

\textit{2. Sublimation of icy volatiles and ice-dust aggregates.} The target nuclei of comet 103P/Hartley 2, spotlighted by its extended mission EPOXI in 2010 \citep{A'Hearn et al.2011}, provides us reliable evidence for the existence of icy grains and aggregates in the coma. In addition, many studies suggest that the polarization characteristics in the inner coma ($\le$ 1000 -- 2000 km) would be induced by the particles whose scattering properties are different from those of the background coma dust, particularly by icy particles \citep[e.g.,][]{Levasseur-Regourd et al.1999,Hines et al.2014}. According to \citet{Mukai1986}, the sublimation lifetime of 1--10$^3$ \micron\ icy grains with absorbing inclusions at 1 au is extremely short ($<$ a few hours), but that of pure water ice at 1 au is nearly a day ($\sim$ $10^{\rm 5}$ seconds). Assuming the possible velocity range of icy particles (100 -- 500 m/s), it takes approximately (1--6) $\times$ $10^4$ seconds (i.e., 3--17 hours) to reach the apex of $P_{\rm r}$ at approximately 6000 km from the center of the nucleus. This value is in good agreement with the lifetime of pure water ice and other icy volatiles. \citet{Perovich1998} also reported that dry and rough ices exhibit very low $P_{\rm r}$ values, approximately 5 \% at a wavelength of 700 nm, whose outward sublimation would result in a rapid increase of $P_{\rm r}$. Additional analysis to confirm the existence of any ice sublimation is not possible owing to the broad-band filters we employed. Therefore, contemporaneous narrow-band imaging data and NIR spectroscopic data will be vital for further discussion.

\textit{3. Disaggregation of cometary dust particles.} Cometary dust has been considered to be composed of fluffy aggregates of submicrometer-sized grains \citep{Fulle et al.2000,Kimura et al.2006} and/or of hierarchical aggregates \citep{Bentley et al.2016,Skorov et al.2016}. Owing to the high fragility of aggregates, they are vulnerable to the differential solar radiation pressure \citep{Boehnhardt et al.1990}; consequently, most of them undergo disaggregation after ejection. \citet{Jewitt2004} presented a similar discussion of the circumnucleus halo with a subsequent rapid increase in $P_{\rm r}$ of comet 2P/Encke. \citet{Jewitt2004} paid attention to the redder, less-polarized inner coma than the adjacent area and conjectured that disaggregation of porous aggregates is one of the likely mechanisms underlying features such as that within 10\arcsec~in Figure \ref{fig:radpolgas}-(b). As discussed above, sublimation of an icy volatile matrix would also be able to trigger disaggregation of dust particles.

Our study of C/2013 US10 questions the conventional classification of polarimetric classes of comets in the optical domain by focusing on the importance of correcting for depolarizing gas emission. From the simple manipulation of the multi-band data taken simultaneously with the polarimetric data, we separated the gas and dust signals and showed that the dichotomy of high- and low-polarization groups of comets does not result from the difference of dust physical properties, but from primarily the spatial distribution of the gas-to-total intensity ratio (i.e., the extent of gas contamination in the image). This conclusion may further support the models of the solar system formation in the respect of similarity in origin of comets.

\section{Summary}
In this paper, we conducted a polarimetric study of C/2013 US10 and found the following:

\begin{enumerate}

\item  The gas intensity rates as a fraction of the total intensity are 5--30\% in the $R_{\rm C}$ band and 3--18\% in the $I_{\rm C}$ band).

\item  The $R_{\rm C}$-band polarization degree changed from 8 \% to 13 \% in the image, but after correcting for depolarizing gas emission lines, this variation becomes almost negligible ($\le$3 \%).

\item Such variation primarily depends on the gas intensity rates, thus suggesting that gas depolarization significantly influenced the observed polarization degree.

\item After considering the effect of gas, we derived the polarization degree in the outer coma, 13.8 $\pm$ 1.0 \% in the $R_{\rm C}$ band and 12.5$\pm$1.1 \% in the $I_{\rm C}$ band for dust, whereas it is 1.0$\pm$2.7 \% for gas. The increments of polarization obtained from the gas correction show that the polarimetric properties of the dust in this low-polarization comet are not different from those in high-polarization comets.

\item The $R_{\rm C}$-band polarization degree of the southeast dust tail (14.4 $\pm$ 2.6 \%), which consists of 0.1--1 mm grains, is equivalent to that in the outer coma in which small and large sized dust particles are mixed.

\end{enumerate}

\acknowledgments

This work at Seoul National University was supported by a National Research Foundation of Korea (NRF) grant funded by the Korean Government (MEST) (No. 2012R1A4A1028713, No. 2015R1D1A1A01060025). We appreciate constructive and valuable comments from anonymous reviewer. Y. G. Kwon was supported by the Global Ph.D. Fellowship Program through the National Research Foundation of Korea (NRF) funded by the Ministry of Education (NRF-2015H1A2A1034260). Also, the operations at these three observatories were supported by the Optical and Near-infrared Astronomy Inter-University Cooperation Program from the Ministry of Education, Culture, Sports, Science and Technology of Japan.



\end{document}